\documentclass[letter]{jpsj2}

\catcode`\@=11
\def\simle{\mathrel{\mathpalette\@versim<}}   
\def\simge{\mathrel{\mathpalette\@versim>}}   
\def\@versim#1#2{\lower2.5pt\vbox{\baselineskip0pt \lineskip-.5pt
   \ialign{$\m@th#1\hfil##\hfil$\crcr#2\crcr\sim\crcr}}}

\title{%
Coexistence of Charge Order and Spin-Peierls Lattice Distortion \\
in One-Dimensional Organic Conductors}
\author{%
Makoto \textsc{Kuwabara}, 
Hitoshi \textsc{Seo}$^{1,2}$ and Masao \textsc{Ogata}$^3$}

\inst{%
Molecular Photoscience Research Center (MPRC), 
Kobe University, Hyogo 657-8501, Japan \\
$^{1}$Correlated Electron Research Center (CERC), AIST 
Tsukuba Central 4, Ibaraki 305-8562, Japan \\
$^{2}$Domestic Research Fellow, Japan Society for the Promotion of
Science, Tokyo 102-8471, Japan \\
$^3$Department of Physics, Faculty of Science, University of Tokyo, 
Tokyo 113-0033, Japan}

\recdate{\today}

\abst{%
The electronic properties of quarter-filled organic 
materials showing spin-Peierls transition 
are investigated theoretically. 
By studying the one-dimensional extended Peierls-Hubbard model 
analytically as well as numerically, 
we find that there is a competition between two different 
spin-Peierls states due to the tetramized lattice distortion 
in the strongly correlated regime. 
One is accompanied by lattice dimerization 
which can be interpreted as a spontaneous Mott insulator, 
while the other shows the existence of 
charge order of Wigner crystal-type. 
Results of numerical density matrix renormalization group 
computations on sufficiently large system sizes 
show that the latter is stabilized in the ground state 
when both the on-site and the inter-site 
Coulomb interactions are large. 
}
\kword{organic conductors, spin-Peierls transition, 
charge order, one-dimensional system, 
density matrix renormalization group, 
bosonization}

\begin{document}
\sloppy
\maketitle


Quasi-one-dimensional (1D) organic conductors
exhibit a variety of electronic states with 
spatially inhomogeneous charge, spin, and lattice structures. 
A typical example is the family of TMTSF$_2X$ and 
TMTTF$_2X$~\cite{Jerome}, where $X$ denotes different anions. 
Recently, the existence of charge ordering (CO) has been identified  
in several of such quasi-1D compounds, 
which has renewed interest in these systems. 
It was first found in DI-DCNQI$_2$Ag, 
a member of the $R_1R_2$-DCNQI$_2X$ family, 
with $X$= Ag or Li, and $R_1$ and $R_2$
taking different substitution groups
to modify the DCNQI molecule itself,
in which the charge pattern 
has been identified as a Wigner crystal-type one~\cite{Hiraki,Nogami}. 
Independently, analogous CO 
was predicted theoretically 
to also exist in TMTTF$_2X$ based on the result of 
mean field calculations~\cite{SeoTM},
and soon after it was confirmed experimentally~\cite{NadXF6,Chow}. 

In the two families menioned above, 
the 1D $\pi$-band is quarter-filled 
in terms of electrons or holes.  
The wave vector along the chain direction for the CO state 
mentioned above is 4$k_{\rm F}$, 
corresponding to the period of two molecules, 
which suggests the origin of 
this phenomenon to be the long range nature 
of Coulomb interaction as theoretically 
studied in the past~\cite{earlydays}.
Further studies have been performed 
on explicit models appropriate for the description of
the electronic properties of DCNQI/TMTTF molecules, 
i.e., 1D extended Hubbard models 
of quarter filling with on-site and nearest-neighbor Coulomb
interactions, $U$ and $V$, respectively.~\cite{SeoTM,MilaZotos,Nishimoto,Yoshioka,Tsuchiizu}. 
In those models, 
the CO insulating ground state is actually stable, in general, 
when both $U$ and $V$ are appreciably large compared to the transfer integrals.
\begin{figure}
\begin{center}
\vspace{3mm}
\includegraphics[width=5.5cm]{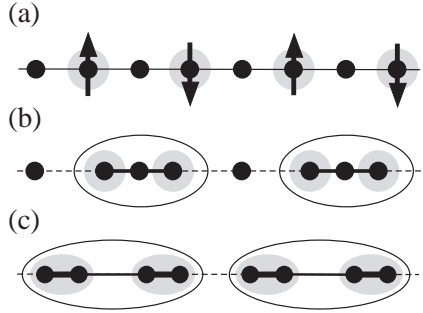}
\caption{Schematic views of the ground states 
in quasi-one-dimensional organic conductors. 
Coexisting states of (a) charge ordering and N{\'e}el ordering, 
(b) charge ordering and spin-Peierls lattice distortion, 
and (c) dimer Mott insulating state 
and spin-Peierls lattice distortion. 
The grey area represents where the charge localizes, 
and the arrows and the ellipses represent 
the ordered spins and the spin singlet formation, respectively. 
The lattice distortions are also shown schematically 
by the thickness of the bonds. 
}
\label{groundstates}
\end{center}
\end{figure}

In the above compounds showing CO,
below the CO transition temperature which is 
typically of the order of 100 K, 
magnetic phase transitions take place, 
to either antiferromagnetic (AF) or spin-Peierls (SP) state. 
These can be understood 
since in this CO state each charge is localized 
on every other site so that  
the spin degree of freedom acts as a quasi-1D spin 1/2 system. 
Then a competition arises between 
the AF state stabilized by the interchain exchange interaction 
and the SP state due to the 
1D instability coupled to the lattice degree of freedom.~\cite{HFSP}
In fact, 
the mean field calculations mentioned above~\cite{SeoTM}, 
expected to be relevant for the former case with 
sufficient three-dimensionality, 
show the coexistence of CO and AF for the ground state 
when $V$ exceeds a critical value. 
This state is schematically shown in Fig. \ref{groundstates} (a), 
which is consistent with the low temperature AF states 
observed in DI-DCNQI$_2$Ag,~\cite{Hiraki} and 
TMTTF$_2X$ with $X=$ Br and SCN~\cite{Jerome,note1}. 

On the other hand, 
in a recent experiment 
on TMTTF$_2X$ with $X=$ PF$_6$ and AsF$_6$,~\cite{Chow} 
the CO is actually found to also persist 
in the SP phase with lattice tetramization. 
Then this coexistence state can be schematically 
drawn as in Fig.~\ref{groundstates}~(b), 
which we call the CO-SP state.  
However, existing theoretical studies, 
mainly numerical ones,~\cite{Riera,Clay} 
have failed to reproduce this CO-SP state 
within the so-called 1D extended Peierls-Hubbard model, 
which is a natural extension of the extended Hubbard model 
to include the electron-lattice coupling. 
In these studies, only one other SP state is found, 
in which the lattice dimerization 
and the lattice tetramization both occur, 
as schematically shown in Fig.~\ref{groundstates}~(c). 
The lattice dimerization produces a spontaneous 
`dimer Mott insulating' (DM) state, 
as will be discussed later, 
so we call this state the DM-SP state~\cite{note2}. 
This state is relevant to 
several DCNQI compounds~\cite{Hiraki,Nogami} 
and to a classical example of SP compounds, MEM-(TCNQ)$_2$~\cite{MEM}. 

In this paper, 
we will show that 
the coexistent state of CO and SP 
is stable in the strongly correlated regime, 
i.e., with large $U$ and $V$, 
in contrast to previous studies.~\cite{Riera,Clay}
This explains naturally the experimental facts noted above, 
and provides a unified view for the ground-state properties 
of the strongly correlated 
quarter-filled 1D organic compounds  
on the basis of the CO phenomenon. 

The 1D extended Peierls-Hubbard model 
as in refs.~\ref{Riera} and \ref{Clay} 
is expressed as 
\begin{eqnarray}
H &=&  \sum_{i,\sigma} t(1+u_i)\left( c^{\dagger}_{i\sigma}
    c_{i+1\sigma} + h.c. \right) 
    + \frac{K}{2} \sum_{i}u_i^2  \nonumber\\
    & & \ \ \ \ \ \ \ \ 
    +U \sum_i n_{i\uparrow}n_{i\downarrow}
    + V\sum_i n_in_{i+1}, 
\label{extPH}
\end{eqnarray}
where $\sigma$ is the spin index which takes $\uparrow$ or $\downarrow$,
$n_{i\sigma}$ and  $c^{\dagger}_{i\sigma}$ ($c_{i\sigma}$) denote
the number operator and the creation (annihilation) operator for the
electron of spin $\sigma$ at the $i$th site, respectively,
and $n_i=n_{i\uparrow}+n_{i\downarrow}$. 
$U$ and $V$ are the on-site and nearest-neighbor Coulomb energies, 
respectively. 
The electron-lattice interaction is included 
in the Peierls-type (or sometimes called the Su-Schrieffer-Heeger-type) 
coupling 
where the intermolecular motions result in the change of the 
transfer integral from $t$ 
to $t(1+u_i)$, 
while it expends the elastic energy $Ku_i^2/2$, 
with $K$ being the spring constant. 
Here $u_i$'s are the normalized displacements
which we treat as classical variables,
and these are to be determined self-consistently to minimize 
the free energy of the system. 

First, let us analyze this model 
within the bosonization scheme~\cite{Emery,FukuTaka} 
following the treatment of Yoshioka {\it et al.}~\cite{Yoshioka}. 
The lattice distortion can be parametrized as 
$u_i= u_d \cos{(\pi x_i/a)}+ u_t \cos{(\pi x_i /2a  + \chi_t)}$~\cite{Riera}, 
where $x_i$ and $a$ are the position of the $i$th site 
and the lattice constant, respectively, 
and $\chi_t$ is a phase factor which is determined in the following. 
We only consider the lattice dimerization 
$u_d$ and lattice tetramization $u_t$, 
which will be confirmed afterwards in the numerical calculations 
that only these modulations are relevant in the ground state. 
The resulting phase Hamiltonian 
for the low-energy properties of this model is given by 
${\cal H}={\cal H}_{\rho} + {\cal H}_{\sigma} 
 + {\cal H}_{1/4} + {\cal H}_{d} + {\cal H}_{t} + {\cal H}_{el}$, 
where 
\begin{equation}
{\cal H}_{\rho}=
\frac{v_{\rho}}{4\pi}
\int {\rm d}x
\left\{
\frac{1}{K_{\rho}} \left( \partial_x \theta_+ \right)^2
+ K_{\rho}\left( \partial_x \theta_- \right)^2
\right\},
\end{equation}
\begin{equation}
{\cal H}_{\sigma}=
\frac{v_{\sigma}}{4\pi}
\int {\rm d}x
\left\{
\frac{1}{K_{\sigma}} \left( \partial_x \phi_+ \right)^2
+ K_{\sigma}\left( \partial_x \phi_- \right)^2
\right\},
\end{equation}
\begin{equation}
{\cal H}_{1/4}= 
g_{1/4}
\int {\rm d}x 
\cos{4 \theta_+},
\end{equation}
\begin{equation}
{\cal H}_{d} =
- g_{d} \ u_d
\int {\rm d}x
\sin{2 \theta_+},
\label{Hd}
\end{equation}
\begin{equation}
{\cal H}_{t} =
- g_{t} \ u_t
\int {\rm d}x
\cos{ \left( \theta_+ - \chi_t - \frac{\pi}{4} \right) }
\cos{\phi_+},
\end{equation}
%
%
and ${\cal H}_{el}$ is the elastic energy term. 
$\theta_+$ and $\phi_+$ are the phase variables for the 
local density fluctuations of the charge and spin with a long wavelengh, 
respectively, 
and $\Pi_\theta = - \partial_x \theta_-/2\pi$ and 
$\Pi_\phi = - \partial_x \phi_-/2\pi$ 
are the corresponding ``momenta'' 
satisfying $[\theta_+(x),\Pi_\theta(x')]=i\delta(x-x')$ 
and $[\phi_+(x),\Pi_\phi(x')]=i\delta(x-x')$. 
Here, we neglected higher order nonlinear terms for the phase
variables as in ref.~\ref{Yoshioka}, 
since these terms have higher scaling dimensions. 

The first two terms, ${\cal H}_\rho$ and ${\cal H}_\sigma$, 
describe the Tomonaga-Luttinger liquid 
where $v_{\rho}, K_{\rho}$ and $v_{\sigma}, K_{\sigma}$
are the so-called Tomonaga-Luttinger parameters 
for the charge and spin parts, respectively. 
The $SU(2)$ symmetry of the nondistorted model with $u_i=0$
requires $K_{\sigma}=1$. 

The next two terms, ${\cal H}_{1/4}$ and ${\cal H}_{d}$, 
are the quarter-filled and the half-filled 
Umklapp scattering terms, 
which are potential energies
in favor of the phase variable to be fixed at $\theta_+=0$ (or $\pi/2$) 
corresponding to the CO state and 
at $\theta_+=\pi/4$ to the DM state, respectively.~\cite{Tsuchiizu} 
The coupling constants  
are calculated in refs.~\ref{Yoshioka} and \ref{Tsuchiizu} 
perturbatively from the weak-coupling regime as 
$g_{1/4}\propto U^2(U-4V)/t^2$ and 
$g_{d}\propto U - AU(U-2V)/t$ where 
$A$ is some numerical constant, 
thus they become prominent in the strongly correlated regime. 
The DM state is analogous 
to that in strongly dimerized quarter-filled organic compounds 
such as $\kappa$-ET$_2X$, 
where the dimerization results in 
an effectively half-filled band~\cite{KanodaKino}. 
However we should note that our DM state 
is possible only when the spontaneous dimerization $u_d$ 
is finite~\cite{Bernasconi}, 
thus it is a consequence of both the strong correlation 
and the 1D instability toward lattice distortion. 

The ${\cal H}_{t}$ term is derived directly from the kinetic energy term 
in the presence of the lattice tetramization $u_t$, 
where the coupling constant $g_{t}$ is proportional to
$t$~\cite{FukuTaka}. 
In the small $U/t$ and $V/t$ limits, 
this term produces 
the conventional weak coupling 2$k_{\rm F}$-CDW state 
due to Peierls instability,  
which fixes the phase variables at 
$\theta_+ = \pi/4, \phi_+=0$, and $\chi_t=0$~\cite{Riera,FukuTaka,note3}. 

Therefore in the case of large $U/t$ and $V/t$ limits, 
which we are interested in, 
there occurs a competition between 
${\cal H}_{1/4}$ and ${\cal H}_d$, 
i.e., between the CO state and the DM state. 
Once the phase variable $\theta_+$ 
is fixed we may substitute its value into ${\cal H}$, 
which result in an effective spin Hamiltonian, 
\begin{equation} 
{\cal H}_{\sigma}'={\cal H}_{\sigma} 
- g_t u_t \int {\rm d}x \cos{\phi_+} 
+ {\cal H}_{el},
\end{equation}
where we have optimized $\chi_t$ so as to gain the energy the most, 
i.e., $\chi_t=\pi/4$ for the CO state 
and $\chi_t=\pi/2$ for the DM state. 
This Hamiltonian is identical to the phase Hamiltonian 
for the SP problem 
in the 1D Heisenberg model, 
where the spin singlet formation due to lattice distortion occurs
even for infinitesimal spin-lattice coupling~\cite{HFSP}.
Thus in our case, 
the lattice tetramization $u_t$ always occurs 
once the electron-lattice coupling 
(and consequently, spin-lattice coupling)
is included, 
then the phase variable of the spin is fixed at 
$\phi_+ = 0$ and the spin gap opens. 
Namely, the CO state and the DM state result in 
the CO-SP state and the DM-SP state, respectively. 
We note that additional 2$k_{\rm F}$-CDW appears 
in the presence of finite $u_t$ 
where the order parameter is proportional to 
$\cos(2k_{\rm F}x_i+\theta_+)\cos{\phi_+}$,~\cite{FukuTaka} 
which is equal to 
$\cos(\pi i /2 )$ and $\cos(\pi i /2 + \pi/4)$ 
for the CO-SP state 
and the DM-SP state, respectively. 

To compare the relative stability of these states, 
the above bosonization procedure is not appropriate, 
not only since the coupling constants are 
obtained perturbatively from the weak coupling regime, 
but also because the treatment of the three nonlinear terms together with 
the lattice degree of freedom is a very subtle problem. 
Instead, we use the numerical density matrix renormalization group (DMRG) 
method~\cite{DMRG} directly to model (\ref{extPH}), 
which is essentially exact. 
The periodic boundary condition is adopted 
and the lattice distortions $u_i$ are calculated self-consistently 
as in ref.~\ref{Riera}. 
We take the number of the states kept in the DMRG method, $m$, up to 250
and check that the $m$-dependence is very small in this region. 

\begin{figure}
\begin{center}
\includegraphics[width=7cm,clip]{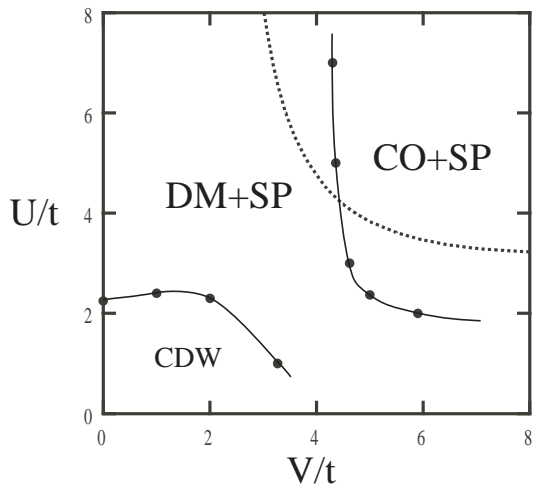}
\caption{Phase diagram of the extended Peierls-Hubbard model 
calculated by the DMRG method in the plane of $U/t$ and $V/t$, 
for fixed values of $1/K=0$ (dotted line) and 1 (filled line).
For definitions of CO-SP, DM-SP, and CDW phases, see text. 
}
\label{phase}
\end{center}
\end{figure}

The obtained results are summarized in Fig.~\ref{phase}, 
which shows the ground-state phase diagram on the plane of 
$U/t$ and $V/t$ for fixed values of $1/K=1$.
For comparison, the phase diagram 
for the case of $1/K=0$ 
corresponding to the purely electronic extended Hubbard model~\cite{MilaZotos} 
is also shown, where there are two phases, 
the Tomonaga-Luttinger liquid metallic phase and the CO
insulating phase.~\cite{MilaZotos,Yoshioka}
The presence of the finite electron-lattice coupling $1/K=1$ makes 
three phases appear, 
the weak coupling CDW state, the DM-SP state, 
and the CO-SP state. 
One can see that 
the CO-SP state is stabilized in a rather wide range of parameters. 
We have noticed that the relative values of the 
ground-state energies for these states, 
and consequently their phase boundaries, 
are rather sensitive to the cluster size $N$ 
when it is small such as $N \simle 20$ as in 
refs. \ref{Riera} and \ref{Clay}. 
Thus we show in Fig.~\ref{phase} the result for $N=36$ 
where such finite size effect is almost negligible, 
which provides qualitatively different results 
from those in the literature, 
especially in terms of the stability of the CO-SP state. 

\begin{figure}
\begin{center}
\includegraphics[width=8cm,clip]{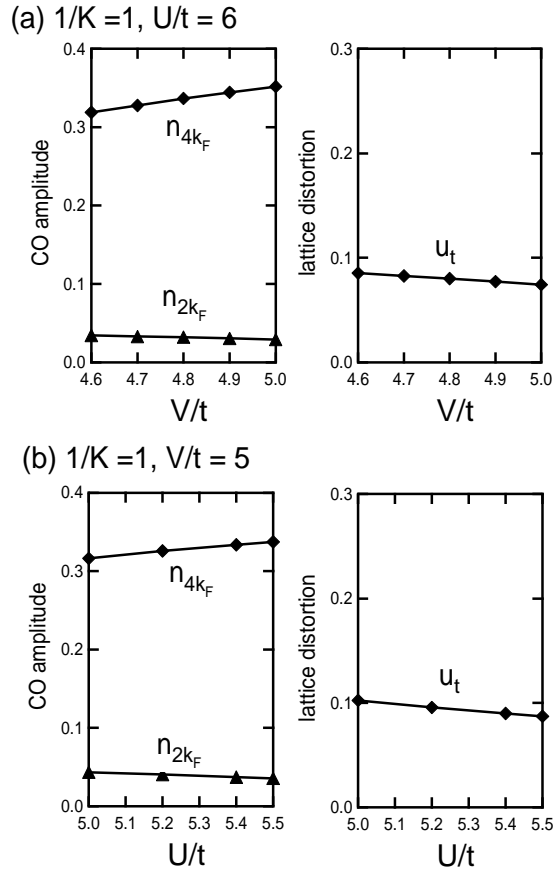}
\caption{Plot of the lattice tetramization 
$u_t$, and the order parameter for 
CO and 2$k_{\rm F}$ CDW, $n_{4k_{\rm F}}$ and $n_{2k_{\rm F}}$, 
respectively, 
for the CO-SP state, 
as a function of (a) $V/t$ 
for a fixed value of $1/K=1$ and $U/t=6$, 
and (b) $U/t$ 
for a fixed value of $1/K=1$ and $V/t=5$.}
\label{plot}
\end{center}
\end{figure}

To observe the property of the CO-SP state in more detail, 
we show the lattice tetramization $u_t$ ($u_d$ is 0), 
and the electron density on each site in the CO-SP state 
as a function of $V/t$ 
for a fixed value of $U/t=6$ 
in Fig.~\ref{plot} (a), 
and as a funcition of $U/t$ for fixed $V/t=5$
in Fig.~\ref{plot} (b), 
both with $1/K=1$. 
The electron density is parametrized as 
$\langle n_i \rangle = 1/2 + n_{4k_{\rm F}} \cos(\pi x_i / a) 
+ n_{2k_{\rm F}} \cos(\pi x_i / 2a ) $, 
where $n_{4k_{\rm F}}$ and $n_{2k_{\rm F}}$ 
are the order parameters for the 4$k_{\rm F}$ CDW, 
i.e., the Wigner crystal-type CO 
and the 2$k_{\rm F}$ CDW, respectively. 
One can see that in the both cases 
of increasing the value of $V/t$ and $U/t$, 
$n_{4k_{\rm F}}$ increases 
while $u_t$ and $n_{2k_{\rm F}}$ decrease. 
These results 
can be interpreted as follows. 
It is natural that the degree of CO 
is enhanced when the 
degree of correlation, $U/t$ or $V/t$ is increased,
thereby increasing $n_{4k_{\rm F}}$. 
In such a case, 
the decrease in $u_t$ and $n_{2k_{\rm F}}$, 
representative of the spin singlet formation due to 
the SP state, 
can be explained by the strong $U/t$ limit, 
where the effective spin exchange is proportional to 
the expectation value for 
both the nearest-neighbor sites 
to be occupied, 
$\langle n_i n_{i+1} \rangle$.~\cite{OgataShiba} 
Then the spin singlet formation energy 
decreases as the tendency toward CO is increased, 
since in the CO state the electron tends to occupy 
every other site. 
This result that $n_{4k_{\rm F}}$ and 
$u_t$ vary in an opposite way 
is consistent with 
the experimental observation that 
the CO and SP states `compete' with each other, 
which is deduced from the variation of their transition temperatures 
as the pressure is applied.~\cite{Chow} 
 

The values of $U/t$ for the actual compounds 
are believed to be approximately $5\sim7$~\cite{Jerome,Hiraki}. 
Therefore the critical value 
of $V/t\sim4$ in our calculations 
necessary to stabilize the CO-SP state is apparently very large. 
Actually, 
a reliable estimate yields $V/t$ of about 2$\sim$3,~\cite{MilaUV} 
which indicates the region of DM-SP in our phase diagram.  
However, this estimated value for $V/t$ is the 
Coulomb interaction between the neighboring molecules 
of the intrachain. 
The interchain Coulomb interaction should also be large, 
since the distance between the chains are rather close, 
especially in TMTTF compounds,~\cite{Jerome} 
although the interchain transfer integrals are small 
due to the anisotropy of the $\pi$-orbital. 
Thus, one should consider 
the parameter $V/t$ 
in our 1D model to be an effective one, 
which is possibly large enough to stabilize the CO-SP state.

Finally, 
let us discuss some effects which we have not included in our model. 
Coupling to anions~\cite{Riera} and/or the so-called 
electron-molecular vibration (e-mv) coupling,~\cite{Clay} 
both of which result in the Holstein-type electron-lattice coupling
modulating the on-site energy, 
help the stabilization of the CO-SP state 
as is studied in refs. \ref{Riera} and \ref{Clay}. 
However, 
these effects should be small 
compared with the Coulomb interactions and 
also with the Peierls-type electron-lattice coupling 
due to the molecular displacements 
which we have studied. 
Therefore we believe that these effects are secondary. 

As for the alternation in the transfer integral 
which exists in TMTTF compounds,~\cite{Jerome} 
as $t(1+u_i) \rightarrow t(1 +(-1)^i \Delta_d)(1+u_i)$, 
where $\Delta_d$ is the degree of alternation, 
we expect it to be slightly disadvantageous 
for the CO-SP state. 
This is because the alternation results in modifying ${\cal H}_d$ 
in Eq.~(\ref{Hd}), 
as $u_d \rightarrow u_d+\Delta_d$ for the first order, 
and the gain in the energy by ${\cal H}$ compared 
to the loss in the lattice elastic energy $\Sigma_i u_d^2$
becomes larger for the DM-SP state. 

On the other hand, 
the value of $1/K$ has been chosen to be fairly large 
in the DMRG calculations, 
in order to stabilize the numerical convergence. 
When we choose realistic values of $1/K$, 
the phase boundary in Fig.~\ref{phase} 
approaches that for $1/K=0$. 
This indicates the stability of the CO-SP state.
However, this again is a subtle problem 
so numerical calculations should be pursued to be conclusive. 

In summary, 
we have argued that the coexistence of charge order and 
spin-Peierls lattice distortion observed in 
quasi-1D organic compounds is 
naturally reproduced by the effects of 
on-site as well as nearest-neighbor Coulomb interaction, 
together with the electron-lattice coupling of Peierls-type. 
Our results show that 
there is a competition between another spin-Peierls state 
coexisting with the spontaneous dimer Mott insulating state 
and that showing the existence of the charge order of Wigner
crystal-type. 
The parameters for the actual compounds 
seems to be in the region near the phase boundary between the two states, 
so that by applying external fields such as pressure 
the ground state is expected to vary, from one to another, 
which is actually observed in TMTTF$_2$AsF$_6$~\cite{Chow}. 

\ 

We thank S. E. Brown, T. Itoh, K. Kanoda, M. Saito, K. Yonemitsu, 
and H. Yoshioka 
for valuable comments and discussion. 
We also appreciate M. Nakamura for kindly providing us with 
the numerical data used in Fig.~\ref{phase}. 


\begin{thebibliography}{99}
\bibitem{note2}In refs. \ref{Riera} and \ref{Clay}, this state is called
	the D$_2$ state and BCDW state, respectively. 
	In addition, we use the terms 
	dimer Mott (DM) insulator and spin-Peierls (SP) distortion 
	for the 4$k_{\rm F}$ and 2$k_{\rm F}$ bond-ordered-wave (BOW),
	often used in the literature, to make their origin more clear.
\bibitem{note3}In ref. \ref{Riera}, this state is called
	the D$_1$ state.
\end{thebibliography}
\end{document}